\documentclass[floatfix,pre,showpacs,preprint]{revtex4-1}

\usepackage{graphicx}
\usepackage{epstopdf}

\usepackage{amsmath}
\usepackage{amssymb}
\usepackage{calc}

\usepackage{bm}
\usepackage[usenames]{color}

\usepackage{latexsym}

\usepackage{verbatim}
\usepackage{xspace}

\def\Lt{L_\text{t}}
\def\rhop{\rho_\text{p}}
\def\sigmap{\sigma_\text{p}}
\def\kt{k_\text{B}T}
\def\mup{\mu_\text{p}}
\def\pp{p} 
\def\phid{\phi_\text{ideal}}
\def\ov{v_\text{OL}}

\def\eg{\textit{e.g.}\xspace}

\begin{document}

\title{Probing a self-assembled \textit{fd} virus membrane with a microtubule}

\author{Sheng Xie and Robert A. Pelcovits}
\affiliation{Department of Physics, Brown University, Providence RI, 02912, U.S.A}
\author{Michael F. Hagan}
\affiliation{Department of Physics, Brandeis University, Waltham MA, 02454, U.S.A.}

\date{\today}

\begin{abstract}
The self-assembly of highly anisotropic colloidal particles leads to a rich variety of morphologies, whose properties are just beginning to be understood. This article uses computer simulations to probe a particle-scale perturbation of a commonly studied colloidal assembly, a monolayer membrane composed of rodlike \textit{fd} viruses in the presence of a polymer depletant. Motivated by experiments currently in progress, we simulate the interaction between a microtubule  and a monolayer membrane as the microtubule ``pokes'' and penetrates the membrane face-on.   Both the viruses and the microtubule are modeled as hard spherocylinders of the same diameter,  while the depletant is modeled using ghost spheres. We find that the force exerted on the microtubule by the membrane is zero either when the microtubule is completely outside the membrane or when it has fully penetrated the membrane. The microtubule is initially repelled by the membrane as it begins to penetrate but experiences an attractive force as it penetrates further.  We assess the roles played by translational and rotational fluctuations of the viruses and the osmotic pressure of the polymer depletant. We find that rotational fluctuations play a more important role than the translational ones. The dependence on the osmotic pressure of the depletant of the width and height of the repulsive barrier and the depth of the attractive potential well is consistent with the assumed depletion-induced attractive interaction between the microtubule and viruses. We discuss the relevance of these studies to the experimental investigations.
\end{abstract}

\pacs{82.70.Dd, 61.30.Cz, 45.50.-j}
\maketitle

\section{INTRODUCTION}
\label{intro}
Colloidal liquid crystals have proven to be a fertile area of experimental and theoretical soft matter research for many years \cite{Dogic2014}.  Of particular interest in recent years \cite{Barry2010, Gibaud2012,Yang2012,Yang2011, Barry2009a, Barry2009, Zakhary2014, Kang2015,Dogic2014,Kaplan2014,Kaplan2013} are assemblies of \textit{fd} viruses in the presence of a polymer depletant which generates an attractive force between the viruses. Each virus is a rod of roughly one micron in length;  the flexibility of the rods can be controlled by molecular engineering and highly monodisperse systems can be fabricated.  As the concentration  of polymer depletant is varied, a variety of equilibrium structures are observed, including membranes, micron-sized monolayer disks, twisted ribbons, braided ribbons, smectic filaments and nematic tactoids. Given that the constituent viruses of these assemblies are micron-sized, it is possible to image these structures at the molecular level. The rodlike virus system has thus served as an important model system to study entropy-driven assembly of hard particles, and has been the subject of numerous experimental and theoretical investigations \cite{Dogic2001,Dogic2003,Kaplan2014}. The assemblies can also be manipulated via optical tweezers achieving structural changes via mechanical means (\eg \cite{Gibaud2012,Sharma2014}. However, the response of such assemblies to particle-scale perturbations has not yet been explored. Here, we computationally consider such a perturbation to the best-studied class of colloidal assemblages, one-rod-length thick monolayers of rodlike particles called colloidal membranes.

The interaction of \textit{fd} viruses in the absence of depletant is well-modeled by the excluded volume interaction of hard rods ~\cite{Purdy2003}. The addition of nonadsorbing depletant introduces an attractive interaction, due to the increased free volume made accessible to the depletant molecules by clustering of the colloidal rods \cite{Asakura1954}. As the polymer concentration is increased (thus increasing the attraction strength), a dilute suspension of virus undergoes a series of phase transitions: from an isotropic phase, to nematic liquid crystalline droplets or tactoids ~\cite{Dogic2001}, to monolayer colloidal membranes \cite{Barry2010}, to a smectic phase consisting of stacks of membranes~\cite{Frenkel2002}. 

The colloidal membranes found at moderate polymer concentrations have been of particular interest as a model system because they exhibit the same long wavelength properties as lipid bilayers, but their micron-scale thickness enables study by light microscopy. Moreover, monolayers of nanoscale rods are of technological importance for the development of scalable optoelectronic devices (\eg \cite{Baker2010,Talapin2004,Zhang2006,Querner2008,Bunge2003}). Colloidal membranes have thus been the subject of experiments studying their continuum-scale properties, morphological transitions induced by optical tweezers, and interactions between pairs of colloidal membranes \cite{Barry2010,Gibaud2012,Zakhary2014,Dogic2014,Wei2013,Barry2009,Sharma2014}.
While early theoretical work focused on nematic tactoids (\eg \cite{Trukhina2008,Trukhina2009,Kaznacheev2002,Kaznacheev2003,Prinsen2003,Prinsen2004,Prinsen2004a,Bhattacharjee2008,Dolganov2007,Haseloh2010,Lishchuk2004,Oakes2007,Otten2009,Verhoeff2011a,Verhoeff2011,Verhoeff2009}), colloidal membranes and smectic stacks have been the subject of more recent modeling (\eg  \cite{Yang2011,Yang2012, Savenko2006,Patti2009,Cuetos2010,Cuetos2008,Kaplan2013,Kaplan2014, Barry2009,Pelcovits2009,Kang2015, Kaplan2010,Tu2013,Tu2013a}).
 Most relevant to the current study, Yang et al.\cite{Yang2012} carried out numerical simulations of the \textit{fd}-depletant system, modeling the viruses as hard spherocylinders of diameter $\sigma$ and length $L$. The depletant was represented by the Asakura-Osawa (AO) model \cite{Asakura1958,Asakura1954}, where the polymers are modeled as ghost spheres which freely interpenetrate each other but interact with the rods via an excluded volume interaction. Monte Carlo (MC) simulations performed by Yang et al. predicted that stable monolayer membranes exist above a critical aspect ratio of the rods (whose value depends on the osmotic pressure of the ghost spheres) and below a critical diameter of the spheres,  approximately $1.7 \sigma$ at low osmotic pressure. The latter prediction was found to be in accord with experimental results \cite{Yang2012}.

While these previous studies have generated significant insights about the continuum-scale behaviors of colloidal membranes, their particle-scale mechanics have yet to be completely characterized. One intriguing way to experimentally probe the structure of a monolayer \textit{fd} membrane is to ``poke'' it face-on with a microtubule of roughly the same diameter as the \textit{fd} virus but of greater length \cite{FungPC2015}, and push the microtubule so that it completely penetrates the membrane. The membrane is held sideways by two optical tweezers and two additional optical traps hold two beads attached to the microtubule or flagellum. The force of the membrane on the rod can be measured as a function of the relative positions of the rod and the membrane.

\begin{figure}
\centering
\includegraphics[width = 4.75in]{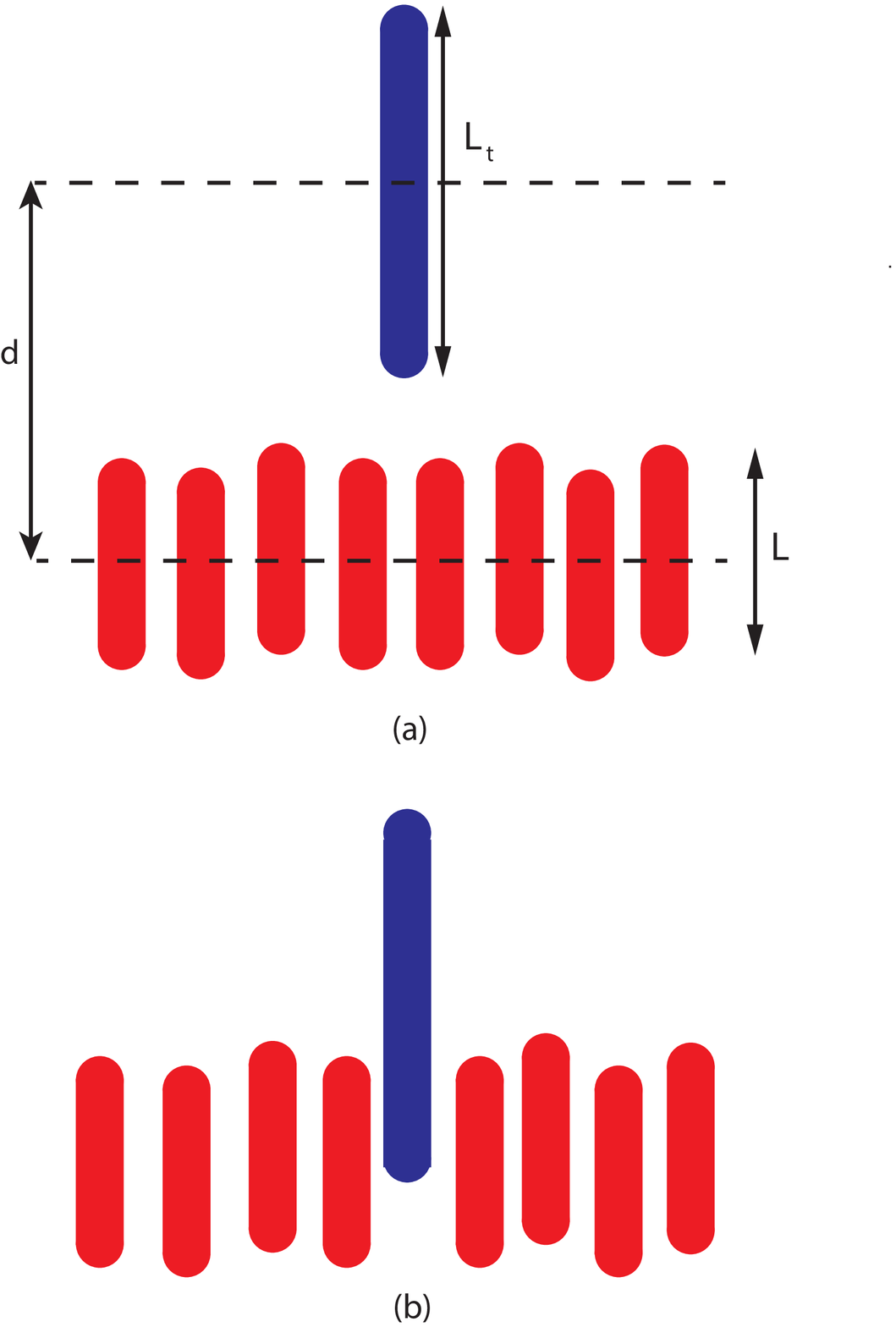}
\caption{(Color online) Schematic illustration of the process modeled in this paper. A microtubule of length $\Lt$ approaches (a) and penetrates (b) a two-dimensional membrane of rods of length $L$. The center of mass separation of the microtubule and membrane is denoted by $d$.
\label{schematic}}
\end{figure}

Motivated by these experiments, which are currently in progress, we consider in this paper a numerical simulation of the interaction potential between a microtubule and virus membrane as a function of their separation. We model the membrane and depletant using the same approach as Ref.~\cite{Yang2012}, and we model the microtubule as an \textit{fd} rod of diameter $\sigma$  but with length $\Lt$, with  $\Lt > L$. Figure \ref{schematic} shows a schematic diagram of the microtubule approaching and penetrating the membrane. To map a phase diagram with computational efficiency, Yang et al.\cite{Yang2012} kept the membrane’s constituent rods fully aligned. Here, we relax that constraint and allow rod rotations, which we find play an important role in the interaction between the membrane and microtubule. Our results provide an experimentally testable prediction for the forces and corresponding free energy profile experienced by a rodlike particle as it enters a colloidal membrane, and enable a first look at the response of an \textit{fd} virus assembly to a particle-scale perturbation.

This paper is organized as follows. In the next section we describe our computational model and the simulation techniques used to measure the interaction potential and force between the microtubule and membrane.  Sec.~\ref{sec:results} presents our results, including a comparison with a theoretical calculation of the potential assuming an idealized configuration of the membrane viruses. We also compare these results with simulations that exclude rod rotations. Sec.~\ref{conclusion} offers concluding remarks.

\section{Simulation Model}

As in Ref.\cite{Yang2012}, we model the the \textit{fd} viruses as hard spherocylinders of diameter $\sigma$ and length $L$. The polymer depletant is represented as ghost spheres of diameter $\sigmap$,  which freely interpenetrate each other but interact with the rods via an excluded volume interaction. Unlike most simulations in Ref.\cite{Yang2012}, we allow for orientational fluctuations of the virus rods. The microtubule is modeled as a spherocylinder with the same diameter as the viruses but of length $\Lt$, with $\Lt > L$.  The long axis of the microtubule is held fixed parallel the $z$ axis, the normal to the plane of the membrane. The microtubule is moved along this axis, which passes through the center of the membrane.  We report our simulation results with $\sigma$ as the unit of length, $\kt$ as the unit of energy, and $\kt\sigma^{-3}$ as the unit of pressure.

The total number, $N$, of rods in the membrane is fixed and the ghost sphere chemical potential $\mup$ is fixed by coupling to a bath with concentration $\rhop$ through insertion/deletion moves. Since the internal pressure of a membrane should be balanced by the polymer solution
pressure, a constant pressure $p$ is maintained in the $xy$ plane by performing volume-change MC moves, while the box size is fixed in the
$z$ direction. MC moves are accepted or rejected according to the Metropolis criterion; the simulations sample from the $N\mup pT$
ensemble. We set the external pressure equal to the sphere osmotic pressure, $\pp$. Since there are no sphere-sphere interactions in the AO model, the osmotic pressure is given by the  van't Hoff equation $\pp=\kt\rhop$.

Yang et al. \cite{Yang2012} mapped out the phase diagram for the virus-depletant system primarily in the absence of rod orientational fluctuations to increase computational efficiency. They studied the phase diagram as a function of both the rod length $L$ (ranging from 20 to 175) and the ghost sphere diameter $\delta=\sigmap/\sigma$ (ranging from 1.2 to 2.0). Since we find that orientational fluctuations are important for the ``poking'' process studied here (shown below), we include orientational fluctuations in our calculations. Due to the increased computational overhead associated with orientational fluctuations, we consider one rod length, $L=100$ and one ghost sphere diameter, $\delta=1.5$. We set $\Lt=150$ for the length of the microtubule.

For these values of $L$ and $\delta$, our simulations with rod orientational fluctuations yield stable monolayer membranes for pressures in the range $0.06 \lesssim p \lesssim 0.1$, while Yang et al. \cite{Yang2012}, excluding orientational fluctuations, found stable isolated membranes for pressures in the range 0.02-0.08 for the same values of $L$ and $\delta$. At higher pressures a smectic phase is stable, even when orientational fluctuations are included. We assess the relative stabilities of the smectic and isolated membrane phases by performing MC simulations with the system initialized in a double layer. In the isolated membrane phase the two layers separate, while in the smectic phase they remain in contact.  However, isolated membranes are highly metastable even at pressures corresponding to the smectic phase. It is thus possible to study the response of a membrane to poking in this regime by initializing the system with a single layer. Similarly, it is possible to experimentally study such metastable membranes by preparing them under conditions in which isolated membranes are stable, and then increasing the osmotic pressure \textit{in situ} \cite{DogicPersonalCommunication}.

We measure the interaction potential $\phi$ and force $F$ between the microtubule and membrane as functions of the distance $d$ between their centers of mass. We consider values of  $d$ in the range $0 \leq d \leq D$, with $D > (L+\Lt)/2$ to study the process of the microtubule approaching and penetrating the membrane.  For our fixed values of $L$ and $\Lt$ we choose $D =150$. Using umbrella sampling \cite{Torrie1977,Frenkel2002a}, we divide the range  $0 \leq d \leq D$ into $n$ partially overlapping windows, each of unit width, except in the range $124.5 \leq d \leq 125.5$ where the microtubule is approaching and penetrating the membrane. In the latter range we use windows of width 0.1.  The overlap between neighboring windows is chosen to be 0.4, except for the windows of width 0.1 where the overlap is 0.04. In total we have $n=(D-1)/0.6 + 1/0.06=265$ windows. In each window, the membrane is initialized in a layer in the $xy$ plane at a high osmotic pressure ($\pp=0.15$) where the layer undulations are smaller. We then adjust $\pp$ to the desired value, and perform MC simulations on this system, measuring the acceptance ratio every 200 cycles, adjusting the maximum amplitudes of the rod translations and rotations and the changes in the $x$ and $y$ dimensions of the simulation box to maintain an acceptance ratio within the range 0.3-0.5. We consider the system to be equilibrated when the all of the amplitudes do not require adjustment for 50 consecutive measurements.

Next, for each window, the microtubule is placed so that its distance $d$ from the membrane lies within the window.  The system is equilibrated via an MC simulation as described above, now including a hard-wall umbrella potential that constrains $d$ to remain within the window, and then simulated for $2\times 10^6$ MC sweeps. The free energy as a function of $d$ is then determined using the weighted histogram analysis method \cite{Kumar1992a}. By equilibrating the system in each window, we are assuming that the microtubule is moving sufficiently slowly so that equilibrium is maintained throughout the process, i.e, the process is reversible. While we present results only for $d>0$, we have also sampled the range $-D \leq d \leq 0$ and found potential plots which are symmetric about $d=0$, up to statistical noise. As one would expect for a reversible process, there is no difference between inserting the microtubule into the membrane and withdrawing it. To improve the statistics of our results we have included data for both positive and negative values of $d$ with the latter data incorporated using the absolute value of $d$. 

\section{Results}
\label{sec:results}

\begin{figure}
\centering
\includegraphics[width = 5.0in]{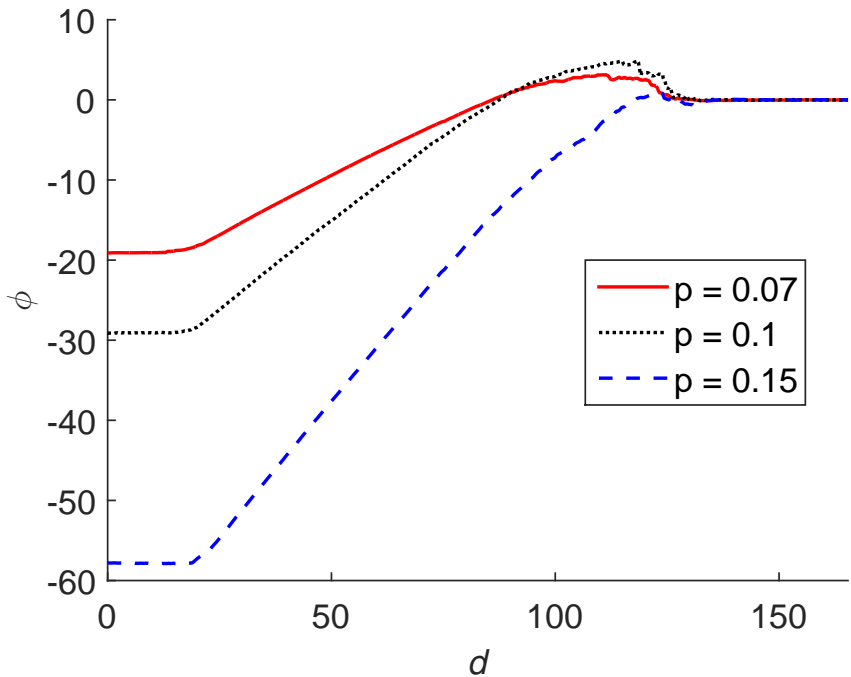}
\caption{(Color online) Simulation results for the interaction potential $\phi$ (in units of $\kt$) of a microtubule ($\Lt = 150$) and a membrane ($N=224,L =100$) as a function of their center of mass separation $d$ for three different pressures: $p =0.07,0.1,0.15$ (in units of  $\kt/\sigma^{-3}$). Lengths are measured in units of $\sigma$. The microtubule fully penetrates the membrane for $d \le 25$.
\label{fig:microtubule-L100}}
\end{figure}

Figure \ref{fig:microtubule-L100} shows the interaction potential $\phi$ as a function of the center of mass separation $d$ of the microtubule ($\Lt=150$) and membrane with $N=224$ rods of length $L =100$ at three different pressures as obtained from the umbrella sampling described in the previous section.  Note that at the highest pressure shown, $p=0.15$, the thermodynamically stable phase is smectic; however, a monolayer membrane can be stabilized by initializing the system of rods in a single layer.

As the pressure is increased there are three changes in the potential profile: the depth of the well increases while both the height and width of the potential barriers decrease. Note that the barriers occur in the vicinity of  $d = 125$ where the microtubule ($\Lt=150$) would enter the membrane ($L=100$) if it were perfectly flat. However, membrane undulations slightly extend the region of interaction between the tube and the membrane, and the span of the barriers is given approximately by: $100 \lesssim d \lesssim130$ for $p=0.07$, $110 \lesssim d \lesssim130$ for $p=0.1$, and $120 \lesssim d \lesssim130$ for $p=0.15$. As we discuss in greater detail below, the changes in the potential profile are consistent with the greater degree of orientational and layer order as the pressure increases and a monolayer of well-aligned rods with minimal undulations or protrusions is formed.

\begin{figure}
\centering
\includegraphics[width = 5.0in]{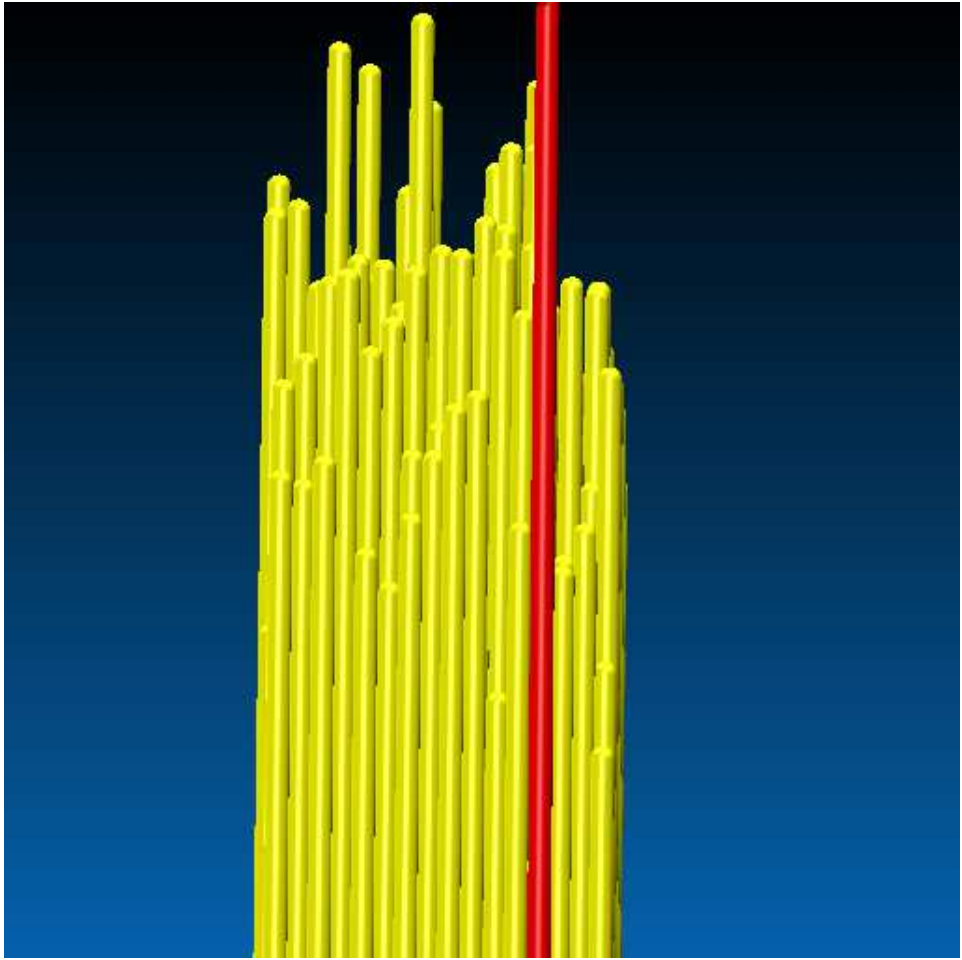}
\protect\caption{ (Color online) Cutaway view of a simulation snapshot of a microtubule (red in color, darker in grayscale) piercing a membrane (yellow or lighter rods) at pressure $p=0.07$. Membrane rods between the observer and the microtubule are rendered invisible to make the microtubule visible. \label{fig:microtubule-piercing-a-1}}
\end{figure}

\begin{figure}
\centering
\includegraphics[width = 5.0in]{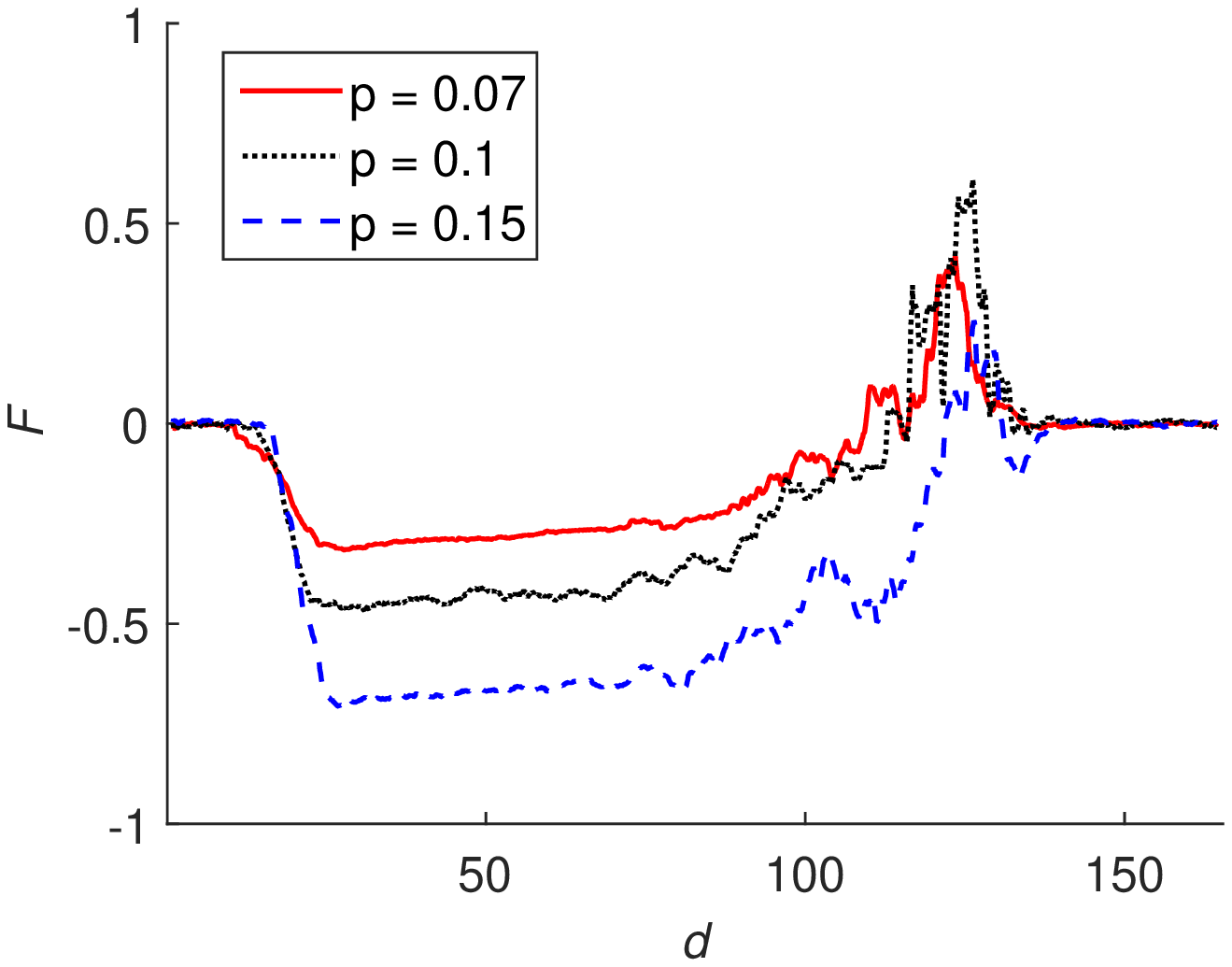}
\protect\caption{(Color online) Force $F$ (in units of $\kt/\sigma$) exerted by the membrane ($N=224,L=100$) on a microtubule ($\Lt=150$) as a function of their center of mass separation $d$ (in units of $\sigma$) for $p=0.07,0.1,0.15$ ( $p$ measured in units of  $\kt/\sigma^{-3}$) as calculated from the potential shown in Fig. \ref{fig:microtubule-L100}. A positive force indicates the microtubule is being pushed out of the membrane.
\label{fig:microtubule-L100-force}}
\end{figure}

The force $F$ exerted on the microtubule by the membrane is given by the derivative of $-\phi$ with respect to $d$. The force corresponding to the potential results of Fig.~\ref{fig:microtubule-L100} is shown in Fig.~\ref{fig:microtubule-L100-force}.   There is a repulsive force ($F>0$) acting on the microtubule as it first penetrates the membrane. As the microtubule penetrates further into the membrane, it experiences an attractive force due to the depletion-induced interaction. The force on the microtubule is zero when it is either completely inside ($|d| \lesssim 25$) or completely outside ($|d| \gtrsim 125$) the membrane.

\begin{figure}
\centering
\includegraphics[width = 5.0in]{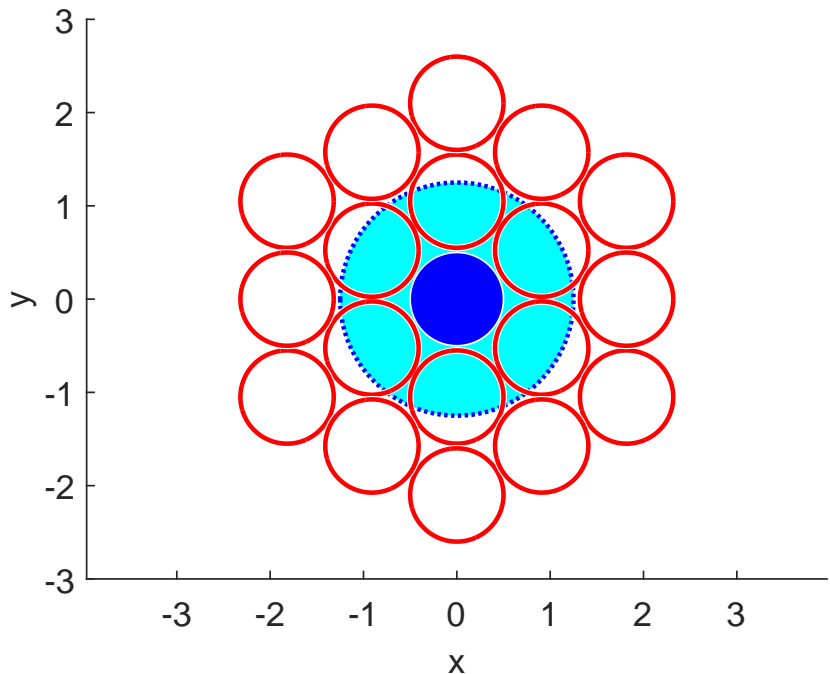}
\protect\caption{(Color online) Idealized membrane configuration used in the theoretical model,  viewed along the $z$ axis, normal to the plane of the membrane. The red circles are the cross section of the virus rods and the blue solid circle at the center of the pattern is the microtubule's cross section. The blue dashed circle
is the outer edge of the depletion zone of the microtubule and has radius $(1 + \delta)/2 = 1.25$ for $\delta = 1.5$, with lengths measured in units of $\sigma$. The overlap volume yielding the depletion potential between the microtubule and membrane is the part of the microtubule's depletion zone that overlaps with any of the rods' depletion zones. \label{fig:Configuration-of-ideal}}
\end{figure}

To investigate the role played by the rotational and translational degrees of freedom of the rods, we first consider an idealized membrane configuration where the depletion interaction between the microtubule and the rods can be calculated exactly. In this simple configuration (Fig.~\ref{fig:Configuration-of-ideal}) the orientation of the rods is fixed parallel to the $z$ axis (thus effects of rod rotations are neglected) and the centers of mass of the rods are fixed on the sites of a hexagonal closed packed lattice in the $xy$ plane. The density of the rods is adjusted to match that of the simulation. Specifically, the densities are 1.0330, 1.0459, 1.0563 for $p=0.07, 0.1, 0.15$, respectively. We use numerical integration to compute the overlap of the excluded volumes of the microtubule and the rods, $\ov(d)$ as a function of $d$. The attractive energy favoring insertion of the microtubule is given by $-\pp \ov(d)$.
We then account for the work performed against the system osmotic pressure in order to create a vacancy for the microtubule to penetrate, by adding $\pp v_\text{rod}$ with $ v_\text{rod}$ the volume per rod in the equilibrated membrane, so the estimated potential is given by $\phid(d)\equiv\pp\left(v_\text{rod}-\ov(d)\right)$. 

\begin{figure}
\centering
\includegraphics[width=3.25in]{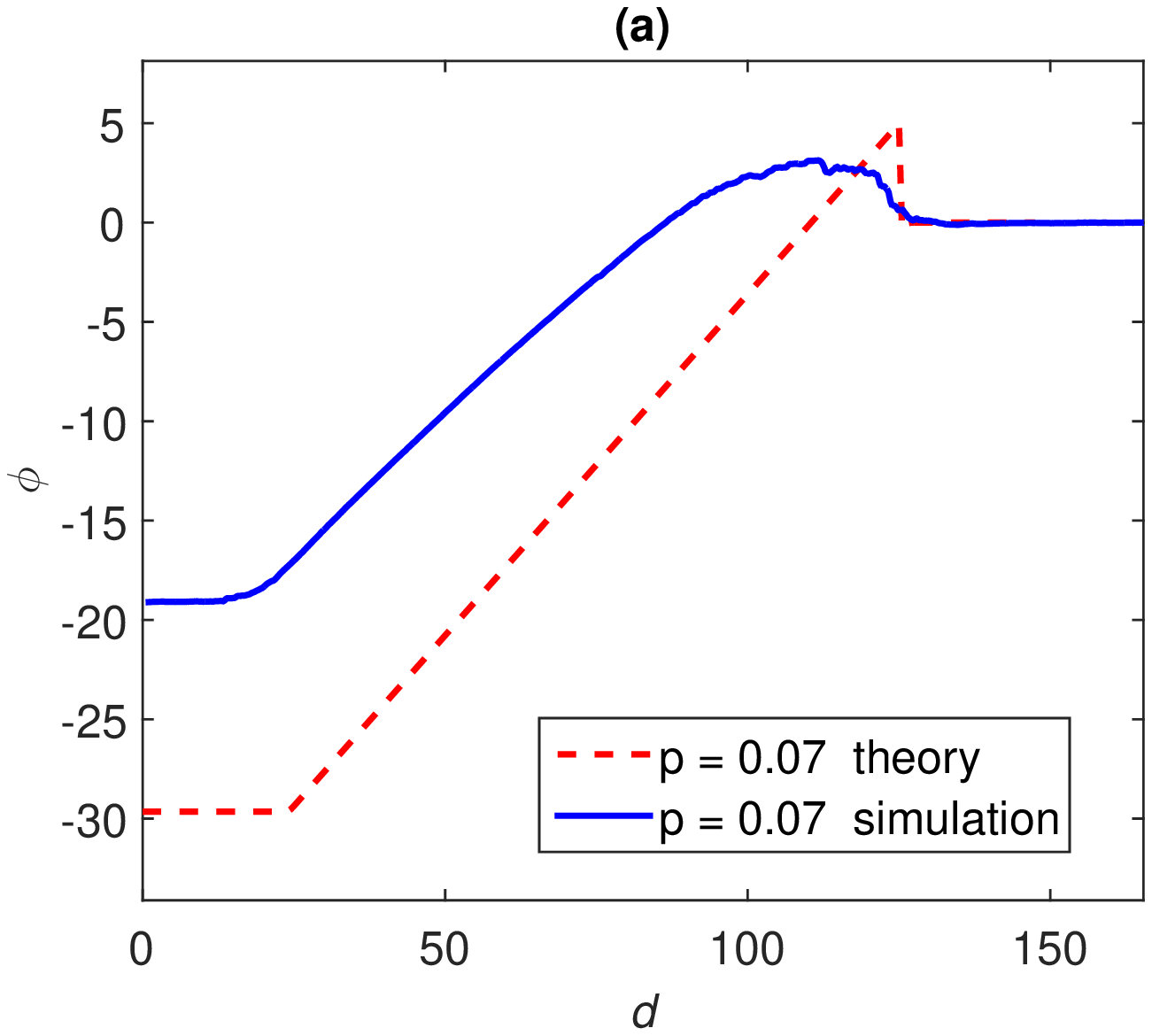}
\includegraphics[width=3.25in]{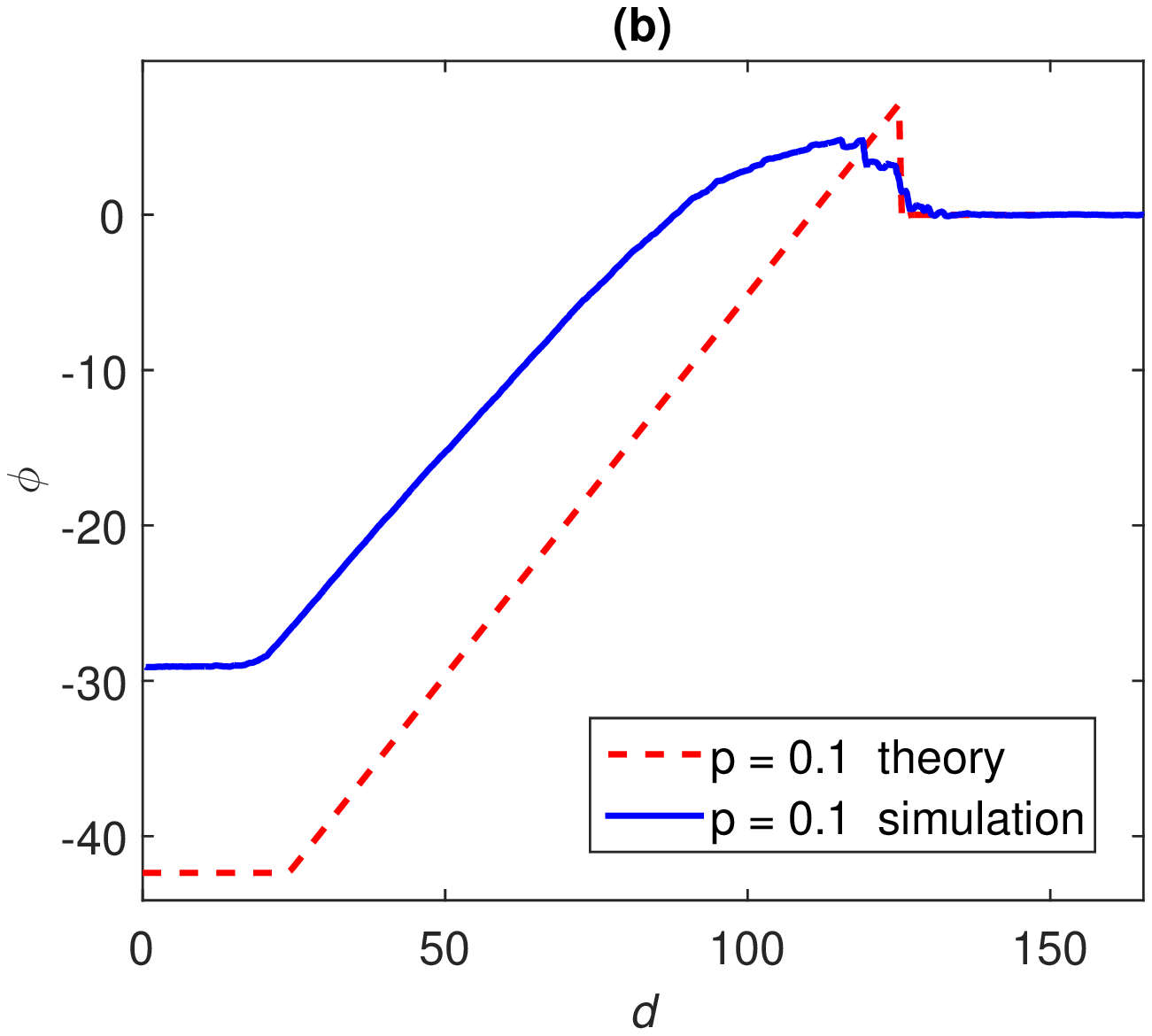}
\includegraphics[width=3.25in]{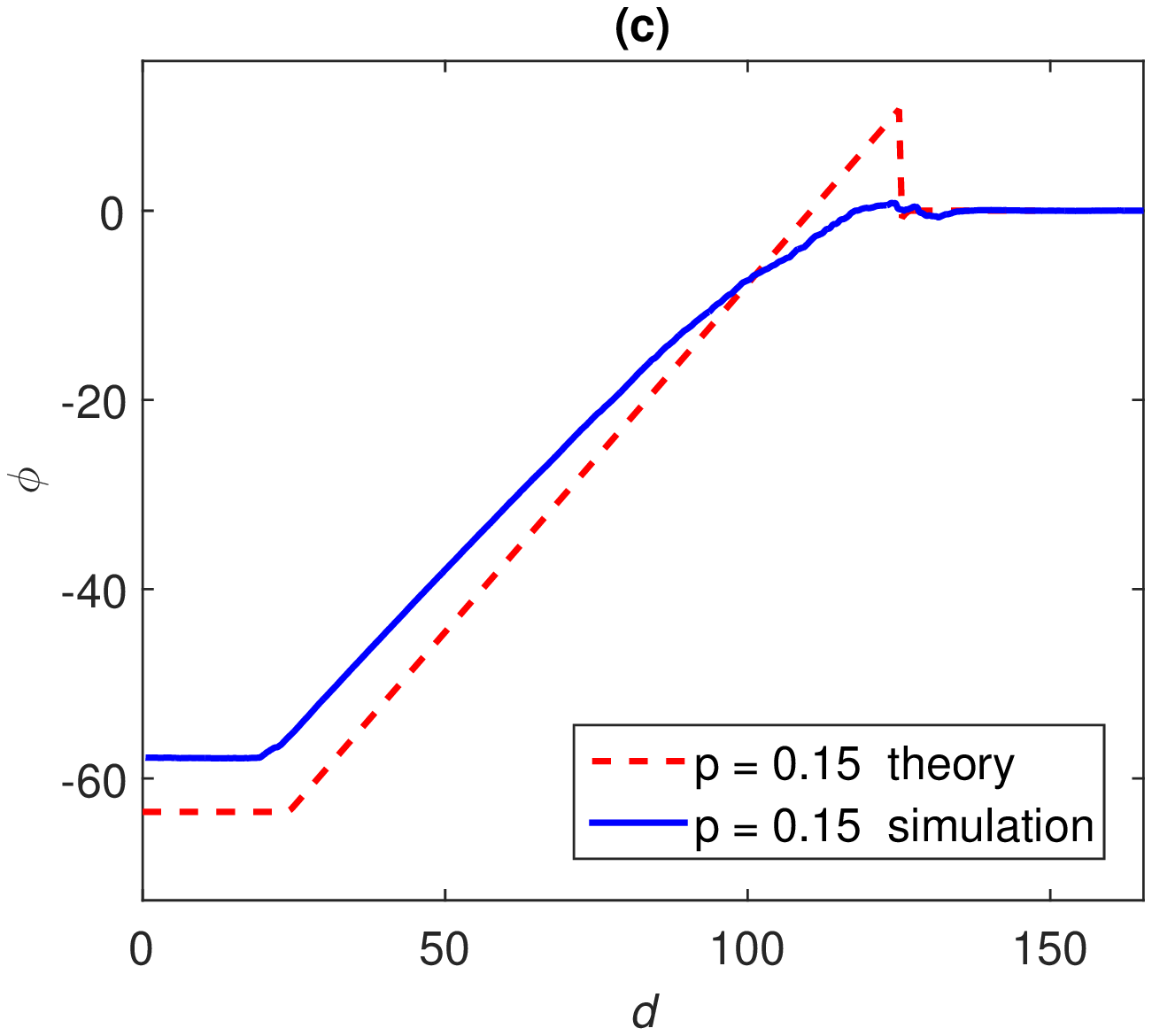}
\protect\caption{(Color online)  Comparison between the simulation results (solid lines) for $\phi$ (Fig.~\ref{fig:microtubule-L100}) and the theoretical excluded volume calculation (dashed lines) of $\phid$ (based on the ideal configuration shown in Fig.~\ref{fig:Configuration-of-ideal}) for a
microtubule ($\Lt=150$) piercing a membrane ($N=224,L=100$) as a function of the separation $d$ for pressures: (a) $p=0.15$; (b) $p=0.1$; (c) $p=0.07$
. The jump in $\phid$ at $d=125$ is the free energy cost of creating the vacancy for the microtubule to penetrate the membrane.
\label{fig:microtubule-compareL100}}
\end{figure}

Figure \ref{fig:microtubule-compareL100} shows $\phi(d)$ at various pressures, comparing the result of the simulations (Fig.~\ref{fig:microtubule-L100}) with the potential $\phid$ calculated on the basis of Fig.~\ref{fig:Configuration-of-ideal}. The jump in $\phid$ at $d=125$ is the free energy cost of creating the vacancy for the microtubule to penetrate the membrane; i.e., the product of the pressure and the rod volume in our idealized model. We observe that as the pressure increases the overall agreement between the simulation and the idealized calculation improves except for the height of the barrier. The improved agreement is physically reasonable, as with increasing pressure the membrane becomes more ordered, with smaller orientational and translational fluctuations (fluctuations which are excluded from the potential $\phid$). Quantitatively, we find that the standard deviations for the distribution of the tilt angle of the rods about the layer normal are $8.8\times 10^{-4}, 7.5\times10^{-4},4.1\times10^{-4}$ at $p=0.07,0.1,0.15$ respectively. The standard deviations for the distribution of the centers of mass of the rods along the normal to the membrane have values 6,4 and 3 at $p=0.07,0.1,0.15$ respectively. However, the simulations show a decrease in barrier height between the lower 2 pressures and the highest value (this is easiest to see in Fig.~\ref{fig:microtubule-L100}), whereas in the calculation of $\phid$  the barrier height is given by the product of the pressure and the rod volume and thus grows monotonically with increasing pressure.

\begin{figure}
\centering
\includegraphics[width=3.25in]{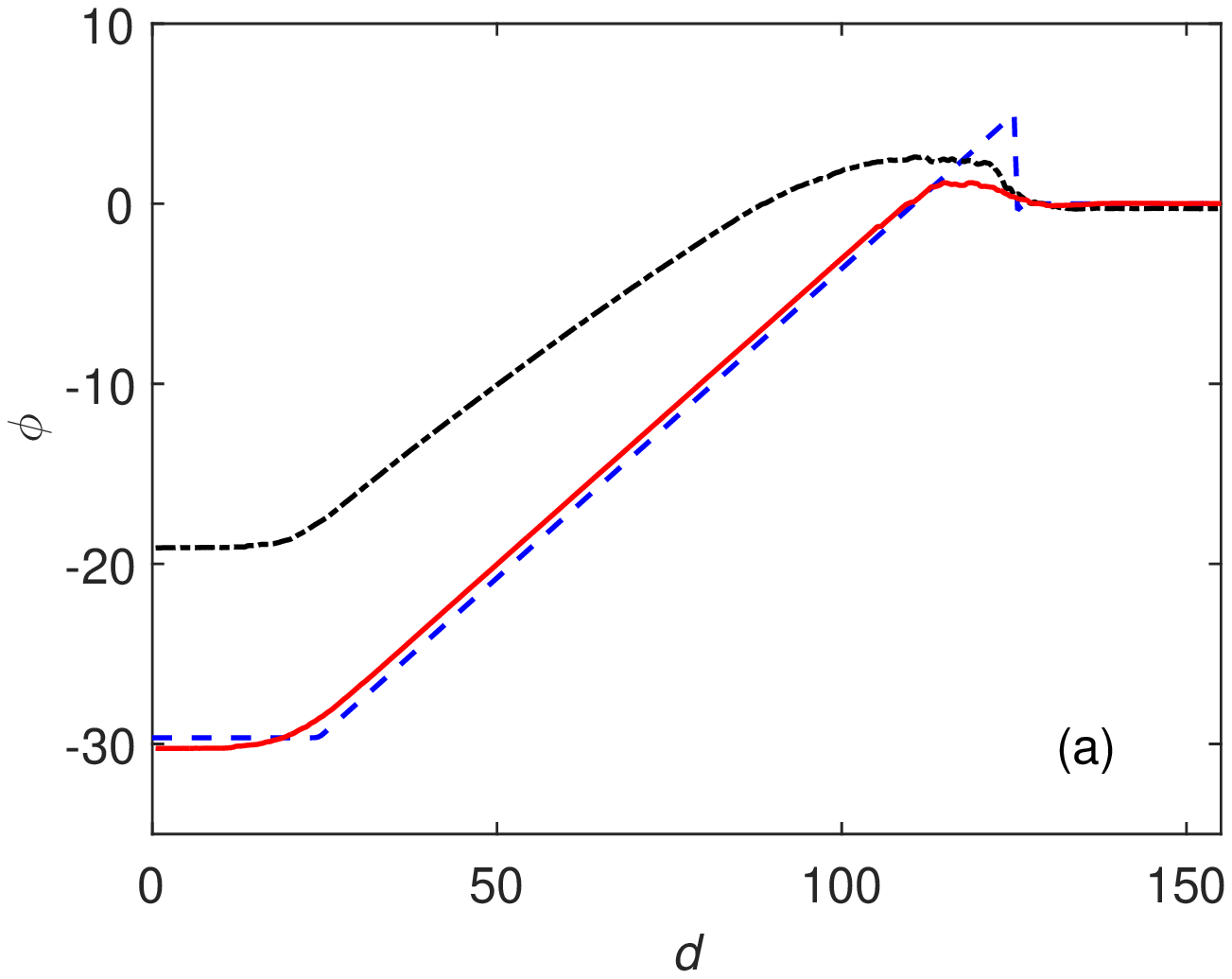}
\includegraphics[width=3.25in]{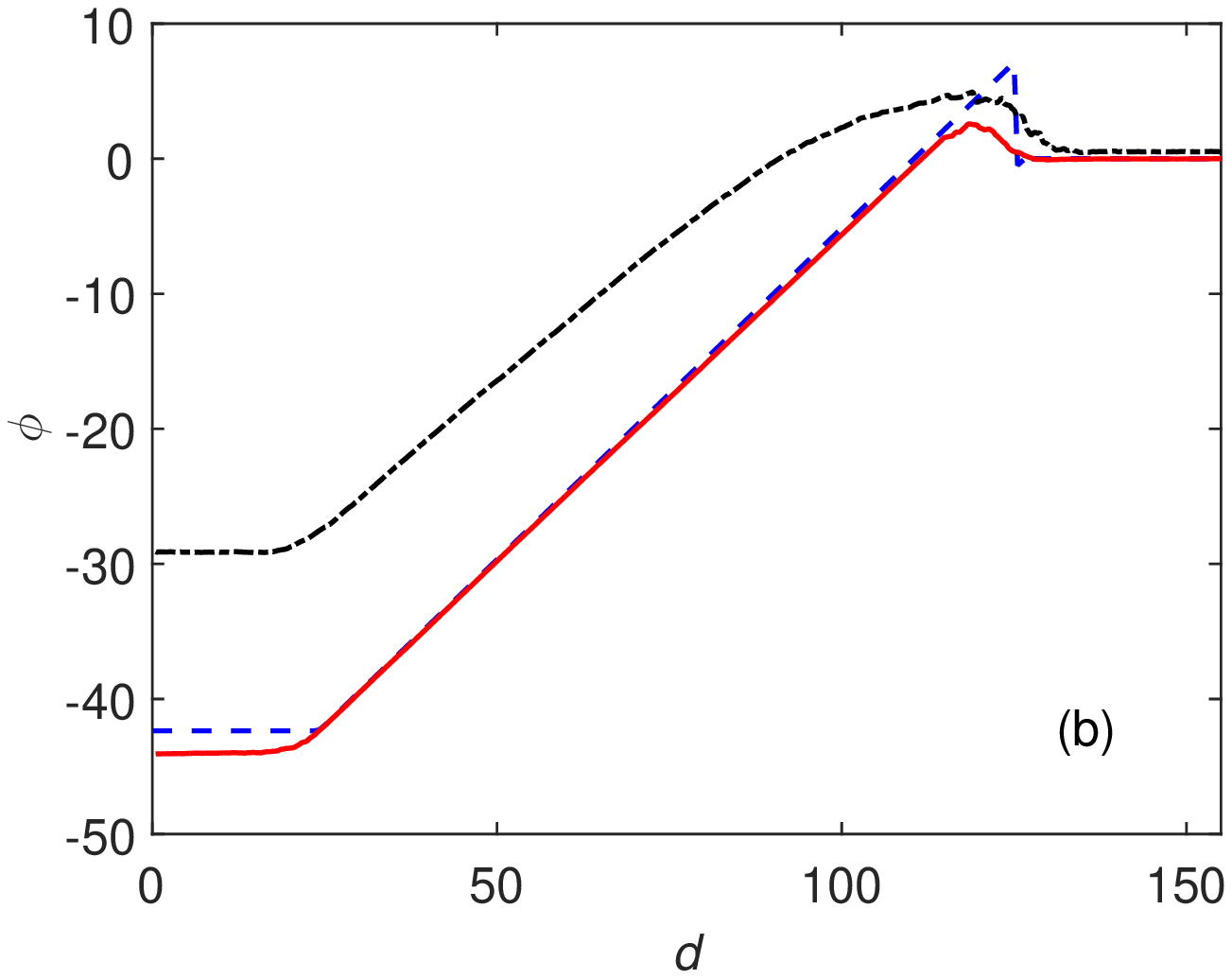}
\includegraphics[width=3.25in]{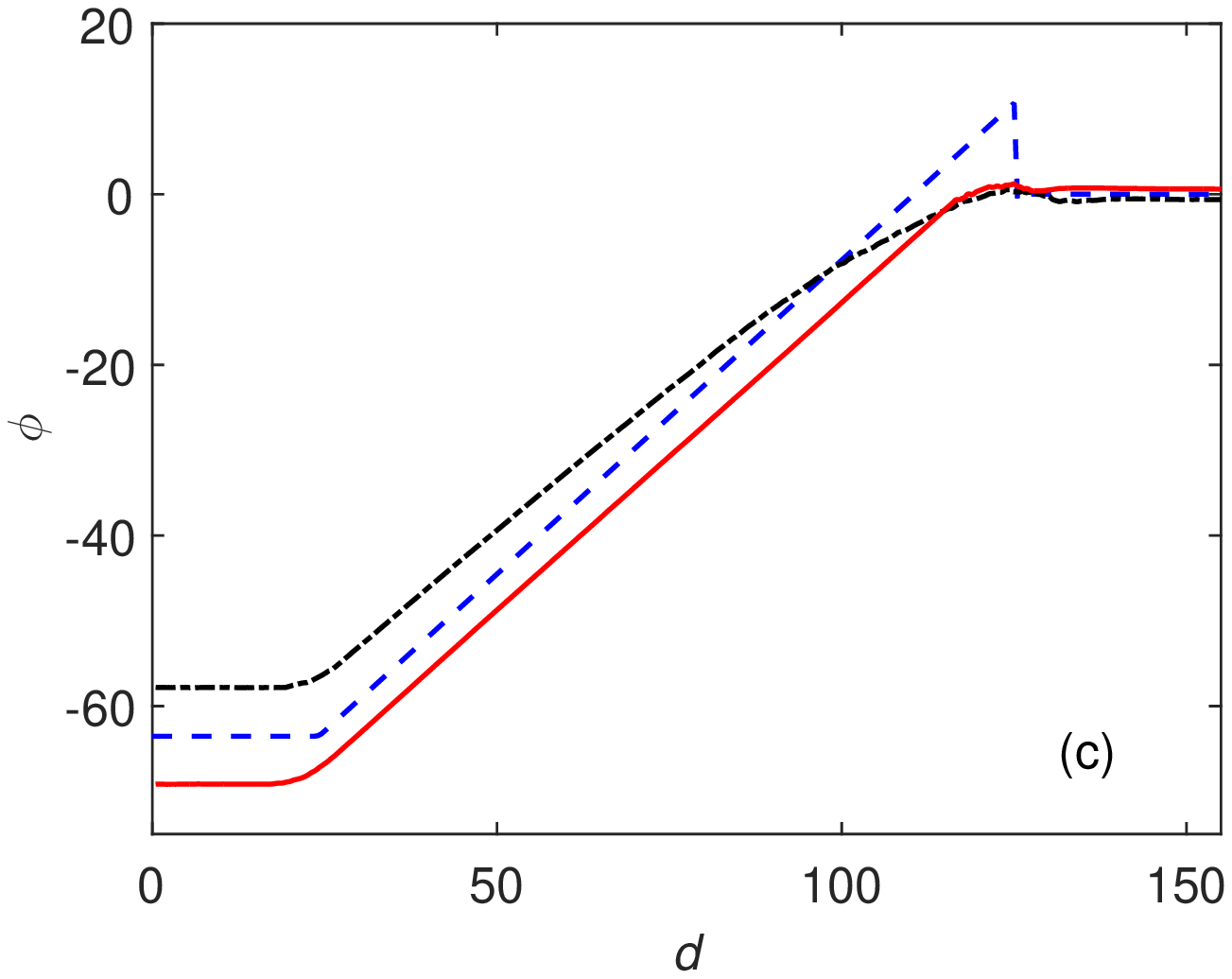}
\protect\caption{(Color online)  Comparison of the results for $\phi$ as a function of the separation $d$ at pressures:  (a) $p=0.07$,  (b) $p=0.1$,  (c) $p=0.15$, obtained from the full simulation  (dotted-dashed black line; Fig.~\ref{fig:microtubule-L100}), the theoretical excluded volume calculation $\phid$  (blue dashed line), and a simulation where rotations of the rods are excluded (solid red line).}
\label{fig:finalcomparison_15}
\end{figure}

To further probe the effects of rod rotations, we perform simulations identical to those described above, but now excluding rotations, while continuing to allow translational motion both within the layer and normal to it (i.e. membrane undulations). The results are shown in Fig.~\ref{fig:finalcomparison_15}, which includes a comparison with $\phid$ and with the results of the original simulations including rotations.  From the figure we see that, except in the barrier regions ($d \simeq 125$), the results of the simulations without rotations are very similar to $\phid$. As the latter does not include any translational motion while the former does, the agreement between the two suggests that the translational degrees of freedom do not significantly affect the potential. As we noted above, the barrier in $\phid$ is computed in a very simplistic fashion, namely the product of the pressure and rod volume, so the disagreement with the simulations, either with or without rotations, is not surprising.

Comparing $\phid$ and the results of the rotationless simulations with the results of the full simulation, we see that the latter exhibits a wider potential barrier, except at the highest pressure where the membrane is well-ordered. In the full simulation the width of the barrier is pressure dependent, less so for the idealized theory and the rotationless simulations.  The depth of the potential well is also smaller in the full simulation even at high pressure (note that the vertical scale in  Fig.~\ref{fig:finalcomparison_15} is larger at higher pressure).  These results are consistent with the nature of the assumed depletion-induced attractive interaction between the microtubule and membrane rods. Rotations of the rods will tend to reduce the overlap of the depletion zones of the microtubules and rod, reducing the depth of the potential well. Orientational fluctuations of the rods also make it more difficult for the microtubule to penetrate the membrane, leading to a taller, wider barrier.

As can be seen in Fig.~\ref{fig:microtubule-L100-force} there is noticeable statistical noise in the force when the microtubule first encounters the membrane. The physical origin of this noise can be understood as follows.  As the microtubule encounters the membrane, local undulations play a critical role in the interaction: the microtubule can be attracted or repelled, depending on the local undulation of the membrane. In particular, the tube is attracted by the depletion force when it is within a depletant-radius of the local membrane surface, but repelled by rod excluded volume as it crosses the local surface. Membrane undulations are not accounted for by our biasing potential, which indexes the distance $d$ between the microtubule and the center of mass of the membrane, i.e., the \textit{average} position of the rods. Therefore, attaining equilibrium from the simulations attaining equilibrium from these situations requires a sufficiently long simulation time such that the tube experiences a representative ensemble of membrane undulations. In principle, greater statistical accuracy could be achieved by including a second potential which either accounts for or biases membrane undulations. Due to finite computational resources, we have not performed such a study.

\section{CONCLUSION}
\label{conclusion}

Using numerical simulations we have studied the penetration of a microtubule into a monolayer membrane of \textit{fd} viruses in the presence of a polymer depletant.  Both the viruses and the microtubule are modeled as hard spherocylinders of the same diameter; the microtubule is 50\% longer than the viruses. The depletant is modeled using ghost spheres \cite{Asakura1958}. The interaction potential and force between the microtubule and membrane as a function of their relative separation $d$ were measured using equilibrium umbrella sampling.  Our results show that the force profile has three distinct regimes:  (1) zero force when the microtubule is completely outside the membrane or when the membrane is fully penetrated, (2) a repulsive force as the microtubule begins to penetrate the membrane, and (3) an attractive force as the microtubule further penetrates the membrane. As the osmotic pressure is increased, the depth of the potential well increases while both the height and width of the potential barrier decrease. These trends are physically reasonable because the membrane is more ordered with increasing pressure.

We have further explored the role played by translational and rotational fluctuations of the rods by computing exactly the interaction potential between the microtubule and a perfectly aligned hexagonal lattice of rods. Excluding the range of $d$ where the microtubule begins to penetrate the membrane (the ``barrier region'') we find that the agreement between the exact calculation and the simulations improves with increasing pressure. We have also performed simulations where we exclude the rotational degrees of freedom of the rods. These simulations agree very well with the exact calculation at all pressures (excluding the barrier region), suggesting that translational degrees of freedom (rod protrusions) do not play a significant role in the interaction potential as compared to the rotational degrees of freedom. This result contrasts with the previous experimental and theoretical observation that rod protrusions play a key role in determining the interaction between vertically adjacent membranes \cite{Yang2011,Yang2012,Barry2010}.

The simulated free energy profiles can be tested against experiments in which optical traps are used to reversibly insert particles into assemblages, such as described in the introduction. Using similar simulation protocols, it would also be possible to determine how insertion forces depend on additional parameters, including the diameter of the inserted particle, finite insertion rates, or alternative assemblage structures. Ultimately, this approach could provide a comprehensive picture of the particle-scale mechanics of colloidal assemblies.

\begin{acknowledgments}

We thank Z. Dogic and J. Fung for helpful discussions.  This work was supported by the NSF through MRSEC Grant No. 1420382.

\end{acknowledgments}


\end{document}